\documentclass{article}
\usepackage{amssymb}

\usepackage{graphicx}
\usepackage{amsmath}

\input{tcilatex}  

\begin{document}

\title{\textbf{Light-front gauge propagator reexamined-II}}
\author{Alfredo T.Suzuki and J.H.O.Sales \\
Instituto de F\'{\i}sica Te\'{o}rica, 01405-900 S\~{a}o Paulo, Brazil.}
\maketitle

\begin{abstract}
Gauge fields are special in the sense that they are invariant under gauge
transformations and \emph{``ipso facto''} they lead to problems when we try
quantizing them straightforwardly. To circumvent this problem we need to
specify a gauge condition to fix the gauge so that the fields that are
connected by gauge invariance are not overcounted in the process of
quantization. The usual way we do this in the light-front is through the
introduction of a Lagrange multiplier, $(n\cdot A)^{2}$, where $n_{\mu }$ is
the external light-like vector, i.e., $n^{2}=0$, and $A_{\mu }$ is the
vector potential. This leads to the usual light-front propagator with all
the ensuing characteristics such as the prominent $(k\cdot n)^{-1}$ pole
which has been the subject of much research. However, it has been for long
recognized that this procedure is incomplete in that there remains a
residual gauge freedom still to be fixed by some ``ad hoc'' prescription,
and this is normally worked out to remedy some unwieldy aspect that emerges
along the way. In this work we propose \emph{two} Lagrange multipliers with
distinct coefficients for the light-front gauge that leads to the correctly
defined propagator with no residual gauge freedom left. This is accomplished
via $(n\cdot A)^2+(\partial \cdot A)^2$ terms in the Lagrangian density.
These lead to a well-defined and exact though Lorentz non invariant
propagator.
\end{abstract}

\section{Introduction}

The history of the light-front gauge goes as far back as 1949 with the
pioneering work of P.A.M.Dirac \cite{dirac}, where the front-form of
relativistic dynamics was introduced as a well-defined possibility for
describing relativistic fields. Since its d\'ebut into quantum field theory
it has known days of both glory and oblivion for varied reasons. On the one
hand it seemed a solid grounded and more convenient approach to studying
quantum fields, e.g., the only setting where a proof of the finiteness of
the $N=4$ supersymmetric Yang-Mills theory could be carried out successfully
was in the light-cone gauge (a facet of its glory) \cite{susylc}. But on the
other hand, manifest Lorentz covariance is lost and non-local terms sneak
into the renormalization program (the other side of the coin that charges us
with a price to pay).

One of the reasons why the light-front form has lured many into this field
of research is due to the fact that its propagator structure seemed simple
enough to deserve their special attention. However, its manifest apparent
simplicity hide many complexities not envisaged at first glance nor
understood without much hard work. For example, one of the, say, ``ugly''
aspects of the ensuing propagator is the emergence of the mistakenly
so-called ``unphysical'' pole which in any physical processes of interest
leads to Feynman integrals bearing these singularities. We say mistakenly
because as it became understood later, it is in fact very much physical in
that without a proper treatment of such a pole, one violates basic physical
principles such as causality \cite{pimentelsuzuki}.

On the other hand, for the brighter side of it, the light-front gauge seemed
advantageous in quantum field theory because it allowed the possibility of
decoupling the ghost fields in the non-Abelian theories, since it is an
axial type gauge, as shown by J. Frenkel \cite{josif}, a property that can
simplify Ward-Takahashi identities \cite{WTI} and problems involving
operator mixing or diagram summation \cite{opmix}.

Looking through the light-front literature we soon realize that there is a
simple and standard gauge vector potential field propagator in which appears
two terms \cite{soper}, namely, 
\begin{equation}  \label{prop}
G^{\mu\nu}_{ab}(k) = \frac {-i\delta ^{ab}}{k^2}\left \{g^{\mu\nu}-\frac{
k^\mu n^\nu+k^\nu n^\mu)}{k\cdot n}\right \}\,,
\end{equation}
where $a, b$ labels non-Abelian gauge group indices.

We see that the propagator (\ref{prop}) has one strictly covariant factor
proportional to the space-time metric $g^{\mu \nu }$ and also the
characteristic light-front factor proportional to $(k^{\mu }n^{\nu }+k^{\nu
}n^{\mu })(k\cdot n)^{-1}$. For the majority of computations, be they in
quantum field theory or in nuclear physics (Bethe-Salpeter, etc.) make use
of this propagator. Some people have recognized the presence of a third term
proportional to $(k^{2}n^{\mu }n^{\nu })(k\cdot n)^{-2}$ \cite{rohrlich}, i.e., 
\begin{equation}
G_{ab}^{\mu \nu }(k)=\frac{-i\delta ^{ab}}{k^{2}}\left\{ g^{\mu \nu }-\frac{
k^{\mu }n^{\nu }+k^{\nu }n^{\mu }}{k\cdot n}+\frac{k^{2}n^{\mu }n^{\nu }}{
(k\cdot n)^{2}}\right\} ,  \label{correct}
\end{equation}
but this third term has always been consistently dropped in the actual
calculations on the grounds that it has been claimed long ago that such 
\emph{``contact terms''} have no physical significance because they do not
propagate any information. After all, from its inception, the paradigm has
always been that gauge terms such as $k^{\mu }n^{\nu }+k^{\nu }n^{\mu }$ and 
$k^{2}n^{\mu }n^{\nu }$ must not contribute to any physical process because
of current conservation. If that be the case, then we must squarely face the
vexing question: Why one would drop only the \emph{``contact terms''} in the
calculations on the grounds that they do not have physical significance
because propagates no information? However, more recently, it has been shown 
\cite{jorgehenrique} that this is not the case. These \emph{``contact terms''} do have physical significance being carriers of relevant information.

Our contribution in this paper is to show that the condition $n\cdot A=0$ ($
n^{2}=0$) is \emph{necessary} but \emph{not sufficient} to define the
light-front gauge. It leads to the standard form of the light-front
propagator (\ref{prop}) which lacks the relevant term of (\ref{correct}).
The \emph{necessary} and \emph{sufficient} condition to uniquely define the
light-front gauge is given by $n\cdot A=\partial \cdot A=0$ so that the
corresponding Lagrange multipliers to be added to the Lagrangian density are
proportional to $(n\cdot A)^2+(\partial \cdot A)^2$ instead of the usual $
(n\cdot A)^{2}$. Note that the condition $\partial \cdot A=0$ in the
light-cone variables defines exactly (for $n\cdot A=A^{+}=0$) the constraint 
\begin{equation}
\label{constraint}
A^{-}=\frac{\partial ^{\perp }A^{\perp }}{\partial ^{+}}\Rightarrow \frac{
k^{\perp }A^{\perp }}{k^{+}}.
\end{equation}
This constraint, together with $A^{+}=0$, once substituted into the
Lagrangian density yields the so-called two-component formalism in the
light-front, where one is left with only physical degrees of freedom, and
Ward-Takahashi identities and multiplicative renormalizability of pure
Yang-Mills field theory is verified \cite{gluonvertex}. Thus, if we start
off by correctly defining the gauge condition in the light-front form, the
problems related to residual gauge freedom, zero modes and ..... are
completely finessed. 

\section{Light-Front Dynamics: Definition}

According to Dirac \cite{dirac} it is \emph{``...the three-dimensional
surface in space-time formed by a plane wave front advancing with the
velocity of light. Such a surface will be called front for brevity ''}. An
example of a light-front is given by the equation $x^{+}=x^{0}+x^{3}$.

A dynamical system is characterized by ten fundamental quantities: energy,
momentum, angular momentum, and boost. In the conventional Hamiltonian form
of dynamics one works with dynamical variables referring to physical
conditions at some instant of time, the simplest instant being given by $
x^{0}=0 $. Dirac found that other forms of relativistic dynamics variables
refer to physical conditions on a front $x^{+}=0$. The resulting dynamics is
called light-front dynamics, which Dirac called front-form for brevity.

A perusal into the specific literature will soon help us to discover that
many different names are used to describe this form of dynamics and the
corresponding gauge, such as light-front field theory, field theory in the
infinite momentum frame, null plane field theory and light-cone field
theory. We prefer the word light-front since the quantization surface is a
light-front (tangential to the light cone).

The variables $x^{+}=x^{0}+x^{3}$ and $x^{-}=x^{0}-x^{3}$ are called
light-front \emph{``time''} and \emph{longitudinal space} variables
respectively. Transverse variables are $x^{\perp }=(x^{1},x^{2})$. We call
the reader's attention to the fact that there are many different conventions
used in the literature. Here, we follow the conventions, notations and some
useful relations employed in \cite{hari1}.

By analogy with the light-front space-time variables, we define the
longitudinal momentum $k^{+}=k^{0}+k^{3}$ and light-front \emph{``energy''} $
k^{-}=k^{0}-k^{3}$.

For a free massive particle, the on-shell condition $k^{2}=m^{2}$ leads to $
k^{+}\geq 0$ and the dispersion relation 
\begin{equation}  \label{disprel}
k^{-}=\frac{(k^{\perp })^{2}+m^{2}}{k^{+}}.
\end{equation}

This dispersion relation (\ref{disprel}) is quite remarkable for the
following reasons: \emph{(1)} Even though we have a relativistic dispersion
relation, there is no square root factor. \emph{(2)} The dependence of the
energy $k^{-}$ on the transverse momentum $k^{\perp }$ is just like in the
nonrelativistic relation. \emph{(3)} For $k^{+}$ positive (negative), $k^{-}$
is positive (negative). This fact has several interesting consequences. 
\emph{(4)} The dependence of energy on $k^{\perp }$ and $k^{+}$ is
mutiplicative and large energy can result from large $k^{\perp }$ and/or
small $k^{+}$. This simple observation has drastic consequences for
renormalization aspects \cite{wilson90}

\section{Massless vector field propagator}

In our previous work \cite{reex1}, we showed that a single Lagrangian
multiplier of the form $(n\cdot A)(\partial \cdot A)$ with $n\cdot
A=\partial \cdot A=0$ leads to a propagator in the light-front gauge that
has no residual gauge freedom left. However, it is clear that the constraint 
$(n\cdot A)(\partial \cdot A)=0$ does not uniquely lead to the \emph{
necessary} conditions $n\cdot A=\partial \cdot A=0$, since the constraint is
satisfied even if only one of the factors vanish. In this sequel work we
propose a more general form with two multipliers each with its corresponding
condition so that they are uniquely defined, and show that we arrive at the
same propagator with no residual gauge freedom left.

The Lagrangian density for the vector gauge field (for simplicity we
consider an Abelian case) is given by 
\begin{equation}
\mathcal{L}=-\frac{1}{4}F_{\mu \nu }F^{\mu \nu }-\frac{1}{2\alpha }\left(
n_{\mu }A^{\mu }\right) ^{2}-\frac{1}{2\beta }\left( \partial _{\mu }A^{\mu
}\right) ^{2},  \label{lag}
\end{equation}
where $\alpha $ and $\beta $ are arbitrary constants. Of course, with these
additional gauge breaking terms, the Lagrangian density is no longer gauge
invariant and as such gauge fixing problem in this sense do not exist
anymore. Now, $\partial _{\mu }A^{\mu }$ doesn't need to be zero so that the
Lorenz condition is verified \cite{Gupta}.

Here, instead of going through the canonical procedure of determining the
propagator as done in the previous section, we shall adopt a more head-on,
classical procedure by looking for the inverse operator corresponding to the
differential operator sandwiched between the vector potentials in the
Lagrangian density. For the Abelian gauge field Lagrangian density we have: 
\begin{equation}
\mathcal{L}=-\frac{1}{4}F_{\mu \nu }F^{\mu \nu }-\frac{1}{2\beta }\left(
\partial _{\mu }A^{\mu }\right) ^{2}-\frac{1}{2\alpha }\left( n_{\mu }A^{\mu
}\right) ^{2}=\mathcal{L}_{\text{E}}+\mathcal{L}_{GF}  \label{2}
\end{equation}

By partial integration and considering that terms which bear a total
derivative don't contribute and that surface terms vanish since $
\lim\limits_{x\rightarrow \infty }A^{\mu }(x)=0$, we have 
\begin{equation}
\mathcal{L}_{\text{E}}=\frac{1}{2}A^{\mu }\left( \square g_{\mu \nu
}-\partial _{\mu }\partial _{\nu }\right) A^{\nu }  \label{3}
\end{equation}
and 
\begin{eqnarray}
\mathcal{L}_{GF}&=&-\frac{1}{2\beta }\partial _{\mu }A^{\mu }\partial _{\nu
}A^{\nu }-\frac{1}{2\alpha }n_{\mu }A^{\mu }n_{\nu }A^{\nu }  \notag \\
&=&\frac{1}{2\beta}A^{\mu }\partial _{\mu }\partial _{\nu }A^{\nu }-\frac{1}{
2\alpha }A^{\mu }n_{\mu }n_{\nu }A^{\nu }  \label{4}
\end{eqnarray}
so that 
\begin{equation}
\mathcal{L}=\frac{1}{2}A^{\mu }\left( \square g_{\mu \nu }-\partial _{\mu
}\partial _{\nu }+\frac{1}{\beta }\partial _{\mu }\partial _{\nu }-\frac{1}{
\alpha }n_{\mu }n_{\nu }\right) A^{\nu }  \label{5}
\end{equation}

To find the gauge field propagator we need to find the inverse of the
operator between parenthesis in (\ref{5}). That differential operator in
momentum space is given by: 
\begin{equation}
O_{\mu \nu }=-k^{2}g_{\mu \nu }+k_{\mu }k_{\nu }-\theta k_{\mu }k_{\nu
}-\lambda n_{\mu }n_{\nu }\,,  \label{6}
\end{equation}
where $\theta =\beta ^{-1}$ and $\lambda =\alpha ^{-1}$, so that the
propagator of the field, which we call $G^{\mu \nu }(k)$, must satisfy the
following equation: 
\begin{equation}
O_{\mu \nu }G^{\nu \lambda }\left( k\right) =\delta _{\mu }^{\lambda }
\label{7}
\end{equation}

$G^{\nu \lambda }(k)$ can now be constructed from the most general tensor
structure that can be defined, i.e., all the possible linear combinations of
the tensor elements that composes it \cite{progress}: 
\begin{eqnarray}
G^{\mu \nu }(k) &=&g^{\mu \nu }A+k^{\mu }k^{\nu }B+k^{\mu }n^{\nu
}C+n^{\mu }k^{\nu }D+k^{\mu }m^{\nu }E+ \notag \\ &&+m^{\mu }k^{\nu
}F+n^{\mu }n^{\nu }G+m^{\mu }m^{\nu }H+n^{\mu }m^{\nu}I+m^{\mu }n^{\nu
}J \label{8}
\end{eqnarray}
where $m^\mu$ is the light-like vector dual to the $n^\mu$, and $A$, $B$, $C$
, $D$, $E$, $F$, $G$, $H$, $I$ and $J$ are coefficients that must be
determined in such a way as to satisfy (\ref{7}). Of course, it is
immediately clear that since (\ref{5}) does not contain any external
light-like vector $m_{\mu }$, the coefficients $E=F=H=I=J=0$ straightaway.
Then, we have 
\begin{equation}
A=-(k^{2})^{-1}  \label{9}
\end{equation}
\begin{equation}
(k\cdot n)(1-\theta )G-\theta k^{2}D=0  \label{10}
\end{equation}
\begin{equation}
(-k-\lambda n^{2})G-\lambda (k\cdot n)D-\lambda A=0  \label{11}
\end{equation}
\begin{equation}
-(k^{2}+\lambda n^{2})C-\lambda (k\cdot n)B=0  \label{12}
\end{equation}
\begin{equation}
(1-\theta )A-\theta k^{2}B+(1-\theta )(k\cdot n)C=0  \label{13}
\end{equation}

From (\ref{10}) we have 
\begin{equation}
G=\frac{k^{2}}{(k\cdot n)(\beta -1)}D  \label{14a}
\end{equation}
which inserted into (\ref{11}) yields 
\begin{equation}
D=\frac{-(k\cdot n)(\beta -1)}{(\alpha k^{2}+n^{2})k^{2}+(k\cdot
n)^{2}(\beta -1)}A  \label{15}
\end{equation}

From (\ref{12}) and (\ref{13}) we obtain 
\begin{equation}
B=\frac{-(\alpha k^{2}+n^{2})}{k\cdot n}C  \label{16}
\end{equation}
and 
\begin{equation*}
C=\frac{-(\beta -1)(k\cdot n)}{(\alpha k^{2}+n^{2})k^{2}+(k\cdot
n)^{2}(\beta -1)}A=D
\end{equation*}

We have then,
\begin{eqnarray}
G^{\mu\nu}(k)&=&-\frac{1}{k^2}\left\{g^{\mu\nu}+\frac{(\alpha k^2+n^2)(\beta-1)}{(\alpha k^2+n^2)k^2+(k\cdot n)^2(\beta-1)}k^\mu\,k^\nu \right. \nonumber \\
&&-\frac{(\beta-1)(k^\mu n^\nu+n^\mu k^\nu)}{(\alpha k^2+n^2)k^2+(k\cdot n)^2(\beta-1)}(k\cdot n)\nonumber\\
&&-\left. \frac{1}{(\alpha k^2+n^2)k^2+(k\cdot n)^2(\beta-1)}k^2\,n^\mu n^\nu \right\}
\end{eqnarray}

In the light-font $n^{2}=0$ and taking the limit $\alpha $,$\beta
\rightarrow 0$, we have 
\begin{equation*}
A=\frac{-1}{k^{2}}
\end{equation*}
\begin{equation*}
B=0
\end{equation*}
\begin{equation*}
C=D=\frac{1}{k^{2}(k\cdot n)}
\end{equation*}
\begin{equation*}
G=\frac{-1}{(k\cdot n)^{2}}
\end{equation*}

Therefore, the relevant propagator in the light-front gauge is: 
\begin{equation}
G^{\mu \nu }(k)=-\frac{1}{k^{2}}\left\{ g^{\mu \nu }-\frac{k^{\mu }n^{\nu
}+n^{\mu }k^{\nu }}{k\cdot n}+\frac{n^{\mu }n^{\nu }}{(k\cdot n)^{2}}
k^{2}\right\} \,,  \label{17}
\end{equation}
which has the outstanding third term commonly referred to as \emph{contact
term}, which is exactly the same result as in \cite{prem}.

\section{Conclusions}

We have constructed Lagrange multipliers in the light-front that leads
to a well-defined fixed gauge choice so that no unphysical degrees
of freedom are left. In other words, no residual gauge remains to be
dealt with. Moreover this allows us to get the correct propagator
including the important so-called contact term. As have been proved,
this term is of capital importance in the renormalization of
(Bethe-Salpeter?) ...

We emphasize that in \cite{reex1} the Lagrange multiplier term of the
form $(n\cdot A)(\partial \cdot A)=0$ was such that $n\cdot A=0$ {\em
and} $(\partial \cdot A)=0$ simultaneously. This means that, of
course, $(n\cdot A)+(\partial \cdot A)=0$. Therefore $[(n\cdot
A)+(\partial \cdot A)]^2=0$, or
\begin{eqnarray}
(n\cdot A)^2+(\partial \cdot A)^2+2(n\cdot A)(\partial \cdot A)&=&0
\nonumber \\ (n\cdot A)^2+(\partial \cdot A)^2&=&-2(n\cdot A)(\partial
\cdot A)\,,
\end{eqnarray}
thus establishing the complete equivalence between the two cases. Note
that this equivalence guarantees that we still have decoupling of the
ghost fields from the physical fields.
\vspace{.5cm}

\noindent \textsc{acknowledgements}: A.T.Suzuki wishes to thank CNPq for
partial support under process 303848/2002-2 and J.H.O.Sales is supported by
Fapesp (00/09018-0).

\section{Appendix}

In this Appendix we review basic concepts of gauge invariance, gauge fixing
and gauge choice that are commonly forgotten or taken for granted, but we
deem appropriate to clarify the issues presented in this work. It is clear
that Maxwell's equations 
\begin{equation}  \label{Maxwell}
\partial_\mu F^{\mu\nu}=\partial_\mu\left(\partial^\mu A^\nu-\partial^\nu
A^\mu\right)=0,
\end{equation}
do not completely specify the vector potential $A^\mu(x)$. For, if $A^\mu(x)$
satisfies (\ref{Maxwell}), so does 
\begin{equation}  \label{gauge}
A^{^{\prime}\mu}(x)=A^\mu(x)+\partial ^\mu \Lambda(x),
\end{equation}
for any arbitrary function $\Lambda(x)$. It is also clear that both vector
potentials $A^\mu$ and $A^{^{\prime}\mu}$ yield the same electric and
magnetic fields $\vec{E}(x)$ and $\vec{B}(x)$, which are invariant under the
substitutions 
\begin{eqnarray}
A_0&\rightarrow&A^{^{\prime}}_0=A_0+\partial_0 \Lambda  \notag \\
\vec{A}&\rightarrow&\vec{A}^{\:^{\prime}}=\vec{A}-\vec{\nabla}\Lambda.
\end{eqnarray}
This lack of uniqueness of the vector potential for given electric and
magnetic fields generates difficulties when, for example, we have to perform
functional integrals over the different field configurations. This lack of
uniqueness may be reduced by imposing a further condition on $A^\mu(x)$,
besides those required by Maxwell's equations (\ref{Maxwell}). It is
customary to impose the so-called ``Lorenz condition''\footnote{
This is not a misprint. J.D.Jackson \cite{Lorenz} calls our attention to
this giving first credit to whom it is deserved.} 
\begin{equation}  \label{Lorenz}
\partial_\mu A^\mu(x)=0,
\end{equation}
which is clearly the unique covariant condition that is linear in $A^\mu$.
However, even the imposition of the Lorenz condition does not fix the gauge
potential, since if $A$ and $A^{\prime}$ are related as in (\ref{gauge}),
then both of them will satisfy (\ref{Lorenz}) if 
\begin{equation}
\square \Lambda\equiv \partial_\mu \Lambda^\mu=0.
\end{equation}

When we choose a particular $A^{^{\prime}\mu}$ in (\ref{gauge}), we say that
we have \emph{``fixed the gauge''}. In particular, an $A^\mu$ satisfying (
\ref{Lorenz}) is said \emph{`` to be in the Lorenz gauge''}. Still,
condition (\ref{Lorenz}) does not exhaust our liberty of choice, i.e., it
does not fix completely the $A^\mu$; we can go to the Lorenz gauge from any $
A^\mu$ choosing a convenient $\phi$ such that it obeys 
\begin{equation}
\Box \phi+\partial_\mu A^\mu=0\Rightarrow \partial^{^{\prime}}_\mu A^\mu=0.
\end{equation}

A further transformation 
\begin{equation}
A^{^{\prime\prime}\mu}=A^{^{\prime}\mu}+\partial^\mu \phi^{^{\prime}},
\end{equation}
with $\phi^{^{\prime}}$ obeying 
\begin{equation}
\Box \phi^{^{\prime}}=0,
\end{equation}
will also lead us to $\partial_\mu A^{^{\prime\prime}\mu}=0$. So, a gauge
potential in the \emph{``Lorenz gauge''} will be determined except for a
gradient of an harmonic scalar field. This remnant or residual freedom can
be used to eliminate one of the components of $A^\mu$, such as, for example, 
$A^0$: Choose $\phi^{^{\prime}}$ such that 
\begin{equation}
\partial^0 \phi^{^{\prime}}=-A^{^{\prime}0},
\end{equation}
so that we have $A^{^{\prime\prime}0}=0$ for any space-time point $(t,\vec{x}
)$. Thus, $\partial_0 A^{^{\prime\prime}0}=0$ and the Lorenz condition will
then be 
\begin{equation}
\nabla \cdot \vec{A}=0\;; \qquad A^0=0.
\end{equation}

This gauge is known as the radiation gauge (or Coulomb one, $\nabla \cdot 
\vec{A}=0$). This gauge choice is not covariant, but can be realized in
every inertial reference frame.

This brings us to the analogy in the light-front case: 
\begin{eqnarray}
A^{^{\prime\prime}\mu}&=&A^{^{\prime}\mu}+\partial ^\mu \phi^{^{\prime}}; 
\notag \\
\partial^{+}\phi^{^{\prime}}&=&-A^{^{\prime}+}.
\end{eqnarray}

Therefore, $A^{^{\prime\prime}+}=A^{^{\prime}+}-A^{^{\prime}+}=0$, and we
obtain the following correspondence: 
\begin{eqnarray}
A^0=0 & \longrightarrow & A^+=0;  \notag \\
\nabla \cdot \vec{A}=0 & \longrightarrow & \partial^+ A^--\partial^\perp
A^\perp =0.
\end{eqnarray}

Note that the second equation above is the constraint (\ref{constraint}).
These imply the double Lagrange multipliers (terms for gauge fixing) in the
Lagrangian density herein proposed 
\begin{equation}
\mathcal{L}_{GF}=-\frac{1}{2\alpha}(n \cdot A)^2-\frac{1}{\beta}(\partial
\cdot A)^2.
\end{equation}

\end{document}